\documentclass[english,number]{article}
\usepackage[T1]{fontenc}
\usepackage[latin9]{inputenc}
\usepackage{array}
\usepackage{float}
\usepackage{amstext}
\usepackage{graphicx}
\usepackage[numbers]{natbib}

\makeatletter

\providecommand{\tabularnewline}{\\}
\floatstyle{ruled}
\newfloat{algorithm}{tbp}{loa}
\providecommand{\algorithmname}{Algorithm}
\floatname{algorithm}{\protect\algorithmname}

\newcommand{\lyxaddress}[1]{
	\par {\raggedright #1
	\vspace{1.4em}
	\noindent\par}
}

\usepackage{multicol}
\usepackage[algo2e,vlined,boxed,linesnumbered]{algorithm2e}

\makeatother

\usepackage{babel}
\begin{document}
\title{Parallel and Memory-limited Algorithms for Optimal Task Scheduling
Using a Duplicate-Free State-Space}
\author{Michael Orr \& Oliver Sinnen}
\maketitle

\lyxaddress{\begin{center}
Department of Electrical and Computer Engineering, University of
Auckland, New Zealand
\par\end{center}}
\begin{abstract}
The problem of task scheduling with communication delays is strongly
NP-hard. State-space search algorithms such as A{*} have been shown
to be a promising approach to solving small to medium sized instances
optimally. A recently proposed state-space model for task scheduling,
known as Allocation-Ordering (AO), allows state-space search methods
to be applied without the need for previously necessary duplicate
avoidance mechanisms, and resulted in significantly improved A{*}
performance. The property of a duplicate-free state space also holds
particular promise for memory limited search algorithms, such as depth-first
branch-and-bound (DFBnB), and parallel search algorithms. This paper
investigates and proposes such algorithms for the AO model and, for
comparison, the older Exhaustive List Scheduling (ELS) state-space
model. Our extensive evaluation shows that AO gives a clear advantage
to DFBnB and allows greater scalability for parallel search algorithms. 
\end{abstract}

\section{Introduction}

Efficient schedules are crucial to allowing parallel systems to reach
their maximum potential. This work addresses the classic problem of
task scheduling with communication delays, known as $P|prec,c_{ij}|C_{\mathrm{max}}$
using the $\alpha|\beta|\gamma$ notation \citep{Veltman1990:msc}.
This problem models a program as a directed acyclic graph of tasks,
giving precedence constraints and communication delays, which must
be arranged onto a set of processors in order to produce a schedule
of minimum length. Solving this problem optimally is NP-hard \citep{Sar1989p},
hence this problem is usually handled with heuristic algorithms, of
which a large number have been developed. \citep{HagJan05AHP,YanGer1993LSW,HWACHO89Sch,Sin07TSP,wang2018list}.
The quality of these solutions relative to the optimal cannot be guaranteed,
however. \citep{DM:2009:SPP}.

The complexity of the problem usually discourages attempts at optimal
solving, but possession of an optimal schedule can make an important
difference for time-critical applications. Without optimal solutions,
it is also very difficult to evaluate the quality of approximation
methods. Branch-and-bound algorithms such as A{*} have shown some
promise in finding optimal schedules, with two notable state-space
models having been proposed: exhaustive list scheduling (ELS) \citep{ShaSinAst2010},
and allocation-ordering (AO) \citep{Orr2015:dfs}. AO, the more recent
model, has the advantageous property of being duplicate-free. Under
the ELS model, an A{*} search stores a set of all states encountered
so far, and needs to compare every newly reached state with this set
in order to mitigate the effect of duplicate states. The duplicate-free
property suggests that AO should be more appropriate for the use of
parallel state-search algorithms, as it removes the need for storing
previous states and the related need for synchronisation between the
processors/threads which this entails. It also suggests that AO may
give improved performance for search algorithms that do not traditionally
have the capacity for duplicate avoidance, such as depth-first branch-and-bound
(DFBnB). In this paper, we propose DFBnB of the scheduling problem
under AO, and for comparison under ELS. For both A{*} and DFBnB, we
investigate parallelised shared-memory versions and discuss the design
decisions. All proposed algorithms are experimentally evaluated with
a large set of task graphs of varying properties (e.g. size, structure,
computation-to-communication ration, etc.) and different numbers of
target processors. We vary the number of threads and employ up to
24 cores in the parallel versions. This paper expands on preliminary
work\citep{orr2017further}, with a more extensive evaluation and
additional theoretical discussion throughout.

Section \ref{sec:Background} provides background information on the
task scheduling problem and the state-space models used, and includes
a discussion on related work. Section \ref{sec:Depth-First-Branch-and}
proposes the DFBnB algorithm for the AO model, while Section \ref{sec:Parallel-Search}
investigates the new parallel search algorithms for our scheduling
problem. Both Section \ref{sec:Depth-First-Branch-and} and Section
\ref{sec:Parallel-Search} present the results of the extensive experimental
evaluation. Finally, Section \ref{sec:Conclusions} presents the conclusions
of the paper.

\section{Background and Related Work\label{sec:Background}}

\subsection{Task Scheduling Model}

\begin{figure}
\begin{centering}
\includegraphics[width=2.2cm]{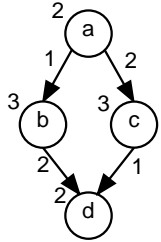}
\par\end{centering}
\caption{\label{fig:A-simple-task}A simple task graph.}
\end{figure}

The problem addressed in this work, $P|prec,c_{ij}|C_{\mathrm{max}}$,
is the scheduling of a task graph $G=\{V,E,w,c\}$ on a set of processors
$P$. $G$ is a directed acyclic graph (or DAG). Each node $n\in V$
is a task belonging to the task graph. Tasks represent an indivisible
block of work that must be performed by a program represented by $G$.
Each task $n_{i}$ has a weight $w(n_{i})$ which represents the computation
time needed to complete that task. An edge $e_{ij}\in E$ represents
that task $n_{j}$ relies on task $n_{i}$; data output from $n_{i}$
is required as an input for $n_{j}$, and therefore $n_{j}$ cannot
begin execution until $n_{i}$ has finished and communicated the necessary
data to $n_{j}$. Each edge $e_{ij}$ has a weight $c(e_{ij})$ which
represents the communication time needed to transmit the necessary
data from $n_{i}$ to $n_{j}$. Figure \ref{fig:A-simple-task} shows
a simple example of a task graph. The target system for our schedule
consists of a finite number of homogeneous dedicated processors, $P$.
Each pair of processors $p_{i,}p_{j}\in P$ possess an identical communication
link. All communication can be performed concurrently and without
contention. Local communication, from $p_{i}$ to $p_{i}$, has no
cost.

\begin{figure}
\begin{centering}
\includegraphics[width=3.1cm]{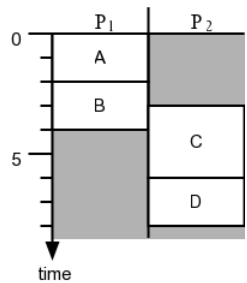}
\par\end{centering}
\caption{\label{fig:A-valid-schedule}A valid schedule for the simple task
graph of Fig.~\ref{fig:A-simple-task}.}
\end{figure}

We aim to produce a schedule $S=\{proc,t_{s}\}$, where $proc(n)$
gives the processor to which $n$ is assigned, and $t_{s}(n)$ gives
the start time of $n$. A valid schedule is one for which all tasks
in $G$ are assigned a processor and a start time, and which satisfies
two conditions for each task. The first condition, known as the processor
constraint, requires that each processor is executing at most one
task at any given time. The second condition, known as the precedence
constraint, requires that a task $n$ may only begin execution once
all of its parents have finished execution, and the necessary data
has been communicated to $proc(n)$. Figure \ref{fig:A-valid-schedule}
shows an example of a valid schedule for the task graph in Figure
\ref{fig:A-simple-task}. An optimal schedule is one for which the
total execution time is the lowest possible.

\subsection{Related Work}

Many combinatorial optimisation techniques could be used in solving
the task scheduling problem. Branch-and-bound search algorithms have
been applied to optimal solving of small problem instances, with some
success \citep{kramer1997branch,fujita2011branch}. Previous work
using the A{*} search algorithm has introduced the ELS \citep{ShaSinAst2010}
and AO \citep{Orr2015:dfs} state-space models, as well as earlier
methods \citep{KwoAhm2005OMT}. A number of pruning techniques and
other optimisations have been developed for ELS \citep{Sinnen2014201},
producing substantial improvements in performance. Many of these techniques
have been adapted for AO as well, and the model has potential for
the development of entirely new pruning techniques. Necessary details
of these two models will be explained in Section \ref{sec:Background}. 

Integer linear programming (ILP) is an alternative combinatorial optimisation
technique which has been applied to this task scheduling problem.
Under ILP, problem instances are formulated as a linear program where
the variables are constrained to integer values. An optimal solution
to the system of equations can then be found, usually with a combination
of standard linear programming algorithms and some variety of branch-and-bound
search. A variety of ILP formulations for the $P|prec,c_{ij}|C_{\mathrm{max}}$
problem have been proposed \citep{el2014new,mallach2016improved,Venugopalan2015:ilp}.
Experiments have shown similarly promising results to those of branch-and-bound,
with neither technique showing a significant advantage over the other
in terms of the size of task graph that can be practically solved.
Popular ILP solvers are mature and highly optimised software packages,
but are generally proprietary. This gives them the benefit of a probable
speed advantage when compared to a custom implementation of state-space
search, but the disadvantage of being somewhat of a ``black box''.
Using a custom implementation makes it easier to understand the behaviour
of the solver, and to gain potentially important insights. It is also
often easier to incorporate domain-specific knowledge into a state-space
model than into an ILP formulation.

Depth-first branch-and-bound has been applied to various NP-hard problems
\citep{zhang1999state,poole2010artificial}. As it requires only a
small and almost fixed amount of memory, it can allow problem instances
to be solved that would be impossible for A{*} using the memory of
a given hardware platform. DFBnB has been previously applied to the
ELS model \citep{venugopalan2016memory}. In this work it was found
that, even when using a pruning technique to avoid a portion of duplicate
states, significantly fewer problem instances could be solved with
DFBnB (or another memory-limited search algorithm, IDA{*}) than with
A{*} while using the ELS model. The lack of duplicates in the AO model
suggests that its use with the DFBnB algorithm will be more successful.

Parallelisation of branch-and-bound has been frequently attempted
\citep{fukunaga2018parallel,ralphs2004library,challou1993parallel,grama1995parallel,burns2010best,jinnai2016abstract},
although no single approach has been found to be dominant over others.
Parallel branch-and-bound has not previously been applied to the problem
of task scheduling with communication delays. 

\subsection{Task Scheduling State-Space Models}

Branch-and-bound is a family of search algorithms commonly used to
solve combinatorial optimisation problems. Through search, they implicitly
enumerate all solutions to a problem, and thereby both find an optimal
solution and prove that it is optimal \citep{BundyWallen1984}. Each
node in the search tree, usually referred to as a state, represents
a partial solution to the problem. Given a partial solution represented
by a state $s$, some operation is applied to produce new partial
solutions which are closer to a complete solution. Performing this
operation to find the children of $s$ is known as branching. Additionally,
every state must be \emph{bounded}: a cost function $f$ is used to
evaluate each state, where $f$ is defined such that $f(s)$ is a
lower bound on the cost of any solution that can be reached from $s$.
These bounds allow the search to be guided away from unpromising solutions,
as a single state can be used to judge the entire subtree that proceeds
from it.

One well known variant of branch-and-bound is a best-first search
method known as A{*} \citep{AICPub834:1968}. A{*} has the major advantage
that it is optimally efficient; using the same cost function $f$,
no search algorithm could find a provably optimal solution while examining
fewer states (disregarding states which have the same $f$-value as
the optimal). A{*} relies on the cost function $f$ to always provide
an underestimate. This means it is guaranteed that $f(s)\leq f^{*}(s)$,
where $f^{*}(s)$ is the true lowest cost of any state in the subtree
proceeding from $s$. A function $f$ with this property is said to
be \emph{admissable. }Algorithm \ref{alg:Pseudocode-A*} gives simple
pseudocode for the A{*}-based scheduling algorithm.

\begin{algorithm}[h]
\begin{algorithm2e}[H]
	\KwIn{$s_{initial}$ is the initial state, an empty allocation}
	\KwIn{$f(s)$, a lower bound on length of schedules reachable from $s$}
	\KwOut{An optimal schedule}	\BlankLine
	$priorityQueue \gets \emptyset$\;
	Insert $s_{initial}$ into $priorityQueue$ with priority $f(s)$\;
	\While{$priorityQueue \neq \emptyset$}{
		$bestState \gets$ pop from $priorityQueue$\;
		\If{$bestState$ represents a complete schedule}{
			\tcp{First complete schedule found must be optimal}
			\Return{$bestState$;}
		}
		\tcp{Solutions that are more complete created by allocating or ordering an additional task}
		\For{$c \in children(bestState)$}{
			Insert $c$ into $priorityQueue$ with priority $f(c)$\;
		}
	}
\end{algorithm2e}

\caption{Pseudocode for the A{*} algorithm.\label{alg:Pseudocode-A*}}
\end{algorithm}

\subsubsection{\label{subsec:Exhaustive-List-Scheduling}Exhaustive List Scheduling
State-Space}

\begin{figure}
\begin{centering}
\includegraphics[width=6cm]{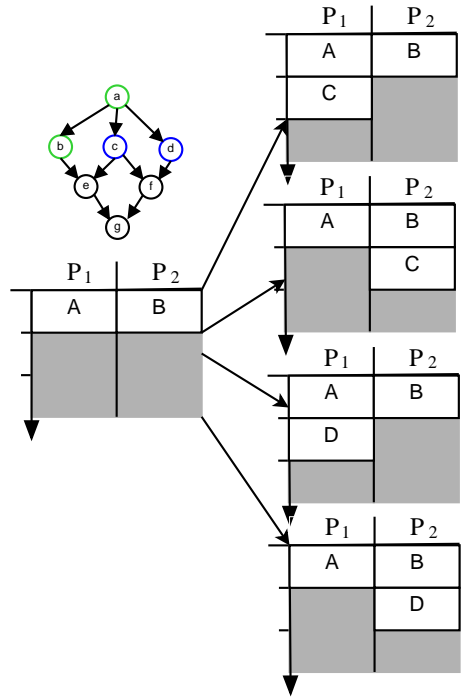}
\par\end{centering}
\caption{Branching in the ELS state space.\label{fig:Branching-ELS}}
\end{figure}

Exhaustive list scheduling (ELS) is a state-space model for optimal
task scheduling which bears a strong resemblance to list scheduling
heuristic methods for approximate task scheduling. In this model,
each state is a partial schedule. Each task is either scheduled or
unscheduled in each state. If a task is scheduled, then it has been
assigned to a processor and given a start time. If a task is unscheduled,
it may be ``free'' or not free. A task is free if all of its dependencies
have been met; that is, if all of its parents have already been scheduled.
At each step, the model branches by placing any free task onto any
processor. The full set of children of a state $s$ therefore represent
all possible free tasks that could have been chosen, and all possible
processors they could have been placed on \citep{ShaSinAst2010}.
In this way, the model simulates all possible decisions that a list
scheduling algorithm could make. An example can be seen in Figure
\ref{fig:Branching-ELS}, demonstrating the four possible child states
of a partial schedule with two processors and two free tasks.

Branch-and-bound methods are most effective when the state-space they
are searching has the property that all subtrees are disjoint. This
means that there is only one path from the root of the tree to any
state in the state-space. When this is not the case, we refer to any
paths that lead to a state already visited 'duplicates'. Equivalently,
any state reached which has already been encountered is termed a 'duplicate'
state. If duplicate states are not detected, then the search algorithm
can perform a substantial amount of unnecessary work: not only might
they examine one duplicate state when an alternate path is found,
they may also proceed to re-explore the entire subtree rooted at that
state. Detecting duplicate states requires keeping a record of those
states which have already been visited, which represents a significant
investment of memory.

\begin{figure}
\begin{centering}
\includegraphics[width=7cm]{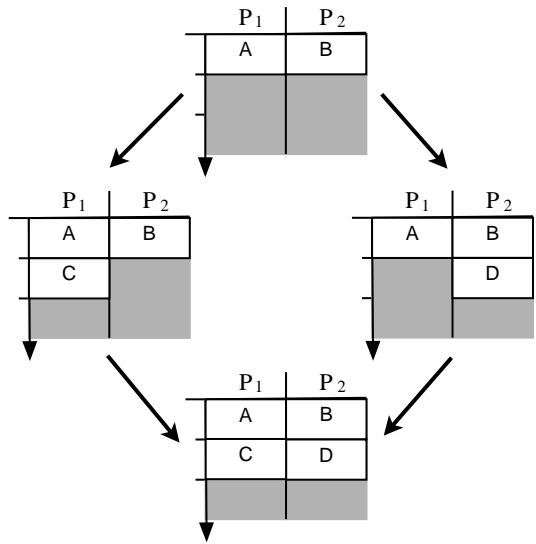}
\par\end{centering}
\caption{Independent decision duplicates.\label{fig:Independent-decision-duplicates.}}
\end{figure}

Unfortunately, the ELS strategy creates a lot of potential for duplicated
states . One type, which we call processor permutation duplicates,
arise when multiple partial schedules can be transformed into each
other merely by relabeling their processors. These duplicates can
be efficiently avoided in ELS using a processor normalisation pruning
technique\citep{Sinnen2014201}. Another type which are fundamental
to the model are independent decision duplicates. Tasks which are
independent of each other can be selected for scheduling in different
orders, but be assigned to the same processors in each case. Performing
the same scheduling decisions in a different order constitutes taking
a different path to reach the same partial schedule, and therefore
a duplicate state. Figure \ref{fig:Independent-decision-duplicates.}
shows two possible paths leading to the same state. These duplicates
can only be avoided by enforcing a specific sequence onto these scheduling
decisions. There is no known method by which this can be achieved
under the ELS model, while still allowing all possible schedules to
be reached.

\subsubsection{Allocation-Ordering State-Space}

Allocation-Ordering (AO) is a new state-space model \citep{Orr2015:dfs}
constructed such that a specific order is enforced on all scheduling
decisions, and therefore the duplicates found in ELS do not exist.
The model is named for the two distinct phases in which it creates
a schedule: first Allocation, where each task is assigned to a processor,
and then Ordering, where a sequence is decided for the set of tasks
allocated to each processor. As it first decides how tasks are grouped
together on processors, this model bears a resemblance to the scheduling
approximation methods known as clustering \citep{gerasoulis1992comparison,kanemitsu2016clustering,liu2011energy},
whereas ELS resembles list scheduling.

In the Allocation phase, a partition of the set of tasks is built
iteratively, with the maximum number of subsets in the partition being
the number of processors available for scheduling. At each step, the
current task can either be added to one of the existing subsets, or
used to begin a new subset. This process allows all possible groupings
of the tasks to be reached, with only one possible path to each grouping.

\begin{figure}
\begin{centering}
\includegraphics[width=6.5cm]{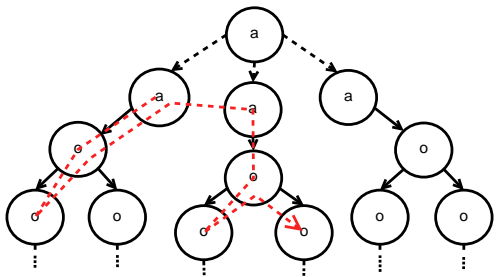}
\par\end{centering}
\caption{A view of the AO state-space, including both phases.\label{fig:Branching-AO}}
\end{figure}

The Ordering phase begins with a complete allocation. From there,
it proceeds in a manner similar to ELS, but on a per-processor basis.
For a processor $p_{i}$, a task $n_{i}$ allocated to $p_{i}$ is
considered to be ``free'' for ordering if there is no task $n_{j}$
also on $p_{i}$ which is an ancestor of $n_{i}$ in the task graph
$G$. At each step, one task is selected from among all those which
are currently free on $p_{i}$, and placed next in order. This is
repeated until all tasks on $p_{i}$ have been given a place in the
sequence. The decision of which processor to order a task on next
can be made arbitrarily, but it must be deterministic such that the
processor which is selected can be determined entirely by the depth
of the current state. A simple round-robin method will suffice. Once
this process has been completed for all processors, a complete schedule
can be derived, simply by giving each task its earliest possible start
time given the processor and place in sequence it has been assigned.
Figure \ref{fig:Branching-AO} provides a view of the overall AO state-space
, showing how the Allocation phase leads into the Ordering phase and
how a search algorithm might move seamlessly back and forth between
them.

By assigning each task to a processor first, and enforcing a strict
order in which the processors are considered, this model eliminates
the possibility of independent decision duplicates. In ELS it was
possible to place task $n_{1}$ on $p_{1}$ and then $n_{2}$ on $p_{2}$,
but equally valid to place $n_{2}$ on $p_{2}$ before placing $n_{1}$on
$p_{1}$. AO can force $p_{1}$ to be considered before $p_{2}$,
making only the first sequence of decisions a possibility.

\section{Depth-First Branch and Bound \label{sec:Depth-First-Branch-and}}

Depth-first branch-and-bound (DFBnB) is a variant of branch-and-bound
which uses a depth-first search strategy, moving as far into the state-space
as possible before back-tracking to try other paths. Just as with
A{*}, a cost function $f$ is used to evaluate each state $s$, producing
a lower bound $f(s)$ on the quality of any solution which could be
reached from $s$. When DFBnB first encounters a state $s_{c}$ representing
a complete solution, the cost $f(s_{c})$ is recorded as $f_{best}$.
Subsequently, if a state is encountered such that $f(s)\geq f_{best}$,
the search will not proceed to that state's children; we have already
found a solution at least as good as any that can be reached from
this state. If another complete solution $s_{c^{\prime}}$ is encountered,
and $f(s_{c^{\prime}})<f_{best}$, then $f_{best}$ is overwritten
with $f(s_{c^{\prime}})$. The search ends when no further states
can be reached with $f(s)<f_{best}$. At this point, the complete
solution with cost $f_{best}$ has been proven to be optimal. Algorithm
\ref{alg:Pseudocode-DFBnB} gives simple pseudocode for the DFBnB
algorithm.

The most obvious advantage of DFBnB when compared with A{*} is its
much lower memory requirements. The best-first nature of A{*} necessitates
the maintenance of a priority queue requiring $O(b^{d})$ space (where
$d$ refers to the depth of the state-space and $b$ is its average
branching factor). A depth-first search requires only states on the
current path and their children to be in memory at any given time,
using $O(bd)$ space. In the case of task scheduling, this means the
memory requirement of A{*} scales exponentially with the number of
tasks, while for DFBnB it scales only linearly. The memory requirement
for DFBnB is small and predictable enough that in practical application
it can usually be treated as constant. A related advantage is that
the data structures used in depth-first search (usually a stack) tend
to be much smaller and simpler, and therefore operations performed
on them are expected to take less time. This is likely to mean that
a depth-first search can process states at a faster rate than a best-first
search.

Naturally, DFBnB also has several disadvantages when compared to A{*}.
The first is that, since it is a depth-first search, it cannot be
applied to state-spaces of infinite depth without careful modification.
As both AO and ELS have a finite (and relatively shallow) depth, this
is not important to our application. 

Another disadvantage is that, unlike A{*}, DFBnB is not optimally
efficient. Like A{*}, DFBnB will always examine every state which
has $f(s)$ less than the optimal solution, but it is likely that
DFBnB will also examine states with $f(s)$ greater than the optimal
solution, which A{*} will never examine. Indeed, the only case in
which DFBnB will not examine extraneous states is if the very first
complete solution it encounters is also optimal. This does not mean
that DFBnB is guaranteed to examine more states in total than A{*};
if A{*} happens to examine a greater proportion of those states where
$f(s)$ is equal to the optimal solution, it can still end up doing
more work. Such situations, however, are strongly implementation-dependent
and unpredictable. It is prudent to assume that, for an arbitrary
problem instance, DFBnB is likely to perform more work overall. 

When applying DFBnB to a state-space containing duplicate states,
there are two possible approaches: ignore the duplicates, or implement
a duplicate-detection mechanism. If we ignore duplicate states, we
are likely to greatly increase the amount of work necessary to find
the optimal solution. A depth-first search will examine every possible
path in the state-space: this could mean that entire sub-trees are
explored many times over. On the other hand, the addition of a duplicate-detection
mechanism will largely negate the main advantage of DFBnB over A{*},
that being its much lower $O(bd)$ memory requirement. In order to
avoid repeating work, the search algorithm must keep a record of states
it has already examined. Although many strategies could exist for
deciding exactly which states should be remembered, any strategy that
is maximally effective at detecting duplicates will require $O(b^{d})$
memory, just as A{*} does. With such an implementation of DFBnB requiring
an exponentially growing amount of memory, and not being optimally
efficient, it is hard to imagine a situation in which it would be
preferable to A{*}.

For those reasons, it seems likely that DFBnB would perform significantly
better on AO, a duplicate-free state-space, than it would on ELS,
a state-space with duplicates. If this is true, then the use of AO
could make DFBnB a more practical option for optimal task scheduling,
making its benefits available for situations where memory is the more
constraining factor.

\begin{algorithm}[h]
\begin{algorithm2e}[H]
	\KwIn{$s_{initial}$ is the initial state, an empty allocation}
	\KwIn{$f(s)$, a lower bound on length of schedules reachable from $s$}
	\KwOut{An optimal schedule}	\BlankLine
	$stack \gets \emptyset$\;
	Push $s_{initial}$ onto $stack$\;
	$bestSolution \gets \infty$\;
	\While{$stack \neq \emptyset$}{
		$currentState \gets$ pop from $stack$\;
		\If{$f(currentState) < f(bestSolution)$}{
			\tcp{Solutions that are more complete created by allocating or ordering an additional task}
			\For{$c \in children(currentState)$}{
				Push $c$ onto $stack$\;
			}
			\If{$currentState$ represents a complete schedule}{
				\tcp{Schedule is best known but not confirmed optimal}
				$bestSolution \gets currentState$\;
			}
		}	}\Return{$bestSolution$;}
\end{algorithm2e}

\caption{Pseudocode for the DFBnB algorithm.\label{alg:Pseudocode-DFBnB}}
\end{algorithm}

\subsection{Evaluation}

To experimentally evaluate the hypothesis that AO would allow better
performance from depth-first branch-and-bound, we performed DFBnB
searches on a set of task graphs using each state-space model. The
evaluation was performed by running branch-and-bound searches on a
diverse set of task graphs using each state-space model. Task graphs
were chosen that differed by the following attributes: graph structure,
the number of tasks, and the communication-to-computation ratio (CCR).
Table \ref{tab:Range-of-task-1} describes the range of attributes
in the data set. A set of 1020 task graphs with unique combinations
of these attributes were selected. These graphs were divided into
three groups according to the number of tasks they contained: 16 tasks,
21 tasks, or 30 tasks. An optimal schedule was attempted for each
task graph using 2, 4, and 8 processors, once each for each state-space
model. This made a total of 3060 problem instances attempted per model.
Gathering this data took approximately one week of continuous server
time.

\begin{table}
\begin{centering}
\begin{tabular}{>{\raggedright}p{3.5cm}|>{\raggedright}p{3cm}|>{\raggedright}p{3cm}}
\textbf{Graph structure} & \textbf{No. of tasks} & \textbf{CCR values}\tabularnewline
\hline 
\begin{itemize}
\item Independent
\item Fork
\item Join
\item Fork-Join
\item Out-Tree
\item In-Tree
\item Pipeline
\item Random
\item Series-Parallel
\end{itemize}
 & \begin{itemize}
\item 16
\item 21
\item 30
\end{itemize}
 & \begin{itemize}
\item 0.1
\item 1
\item 10
\end{itemize}
\tabularnewline
\end{tabular}
\par\end{centering}
\caption{\label{tab:Range-of-task-1}Range of task graphs in the experimental
data set.}
\end{table}
 The algorithms were implemented in the Java programming language.
Existing implementations of both ELS and AO were utilised. All tests
were run on a Linux machine with 4 Intel Xeon E7-4830 v3 @2.1GHz processors.
The tests were single-threaded, so they would only have gained marginal
benefit from the multi-core system. The tests were allowed a time
limit of 2 minutes to complete. For all tests, the JVM was given a
maximum heap size of 96 GB. Each search was started in a new JVM instance,
to remove the possibility of previous searches impacting them through
garbage collection and JIT compilation.

\begin{figure}
\begin{centering}
\includegraphics[width=11cm]{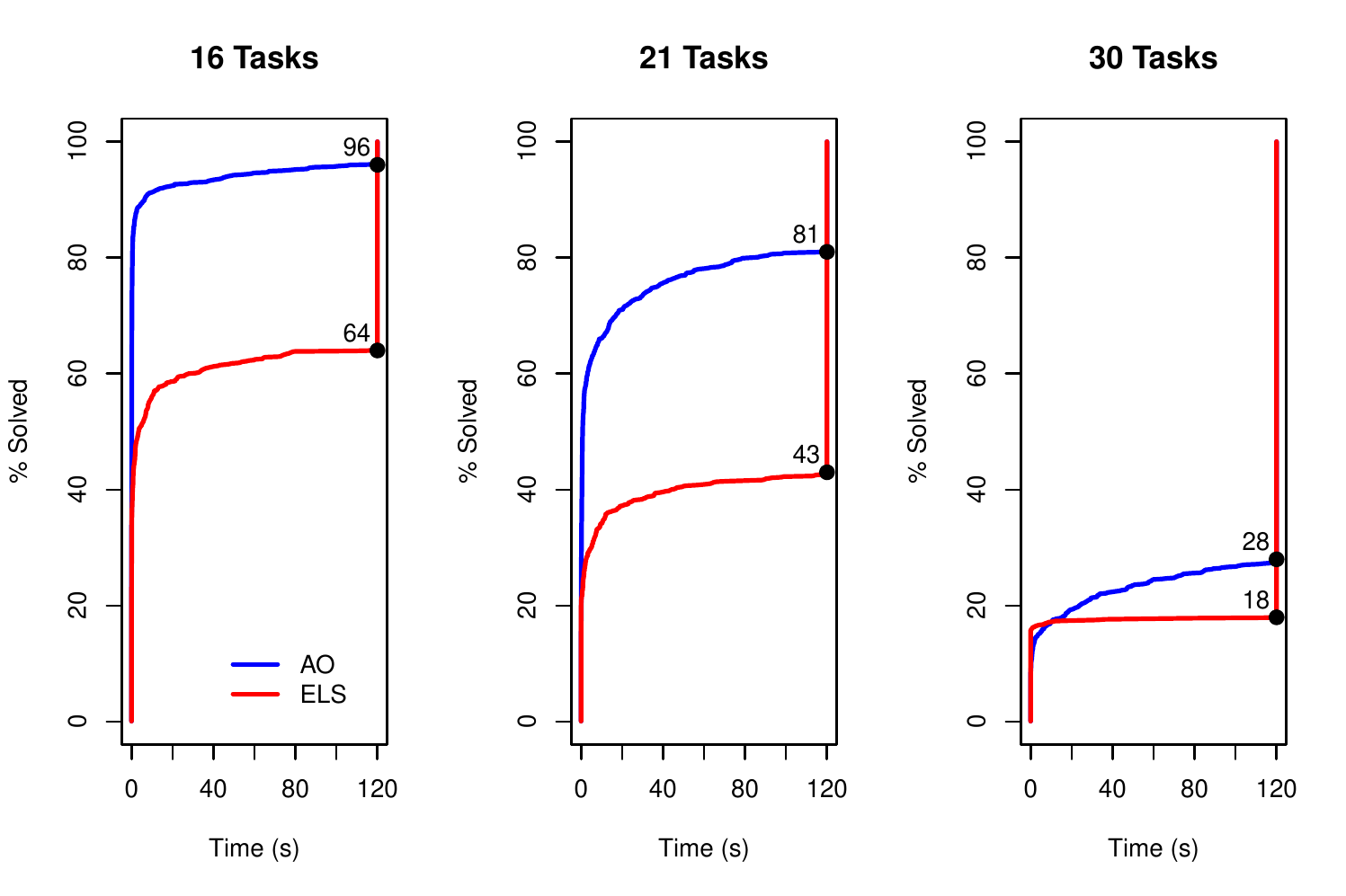}
\par\end{centering}
\caption{Comparing the performance of DFBnB using AO and ELS.\label{fig:dfbnb-comparison}}
\end{figure}

Figure \ref{fig:dfbnb-comparison} shows the results of these tests
as performance profiles: the $x$-axis shows time elapsed, while the
$y$-axis shows the cumulative percentage of problem instances which
were successfully solved by this time. In all three groups of task
graphs, it is clear that substantially more problem instances were
solved with AO than with ELS. In the 16 task group, ELS solved 64\%
of problem instances while AO solved 96\%. This is both a large difference
in absolute terms and a relative advantage of 50\% for AO. Similarly,
in the 30 task group, ELS solved 18\% while AO solved 28\%. This is
a relative advantage of 55\% for AO. In the 21 task group ELS solved
43\% and AO solved 81\%, giving a relative advantage of 88\% for AO.
These results make it clear that AO is a significantly better choice
than ELS when using DFBnB. This performance is comparable to what
was previously observed when using AO with the A{*} algorithm \citep{2019arXiv190106899O},
suggesting that in practice DFBnB does not perform significantly worse
than A{*}. Along with the low memory cost, the performance demonstrated
here makes a strong case for DFBnB as the primary branch-and-bound
method for optimal task scheduling, which is different to previous
results when the ELS model is employed \citep{venugopalan2016memory}.
As well as removing the possibility of failure due to memory exhaustion,
DFBnB could be used to solve many problem instances simultaneously
on a multicore system as there is no competition for memory from which
simultaneous A{*} runs would suffer.

\section{Parallel Search \label{sec:Parallel-Search}}

State-space search is very time consuming, even when using a good
state-space model with effective cost functions and pruning techniques.
Accelerating a search through parallelisation may be critical to obtaining
a solution within an acceptable timeframe. As the AO model is duplicate-free,
it does not require the use of a duplicate-detection mechanism, or
any of the data structures associated with one. In a parallelised
implementation of branch-and-bound, these data structures require
synchronisation between workers, adding greatly to the potential for
contention and likely limiting overall speedup. Therefore, without
the need for duplicate detection, it seems probable that parallel
branch-and-bound could be more effective when used with the AO model
than with the ELS model. In this section we investigate and propose
shared-memory parallel versions of both A{*} and DFBnB. Branch-and-bound
search has several features which make parallelisation a non-trivial
task \citep{talbi2006parallel}. We describe the different factors
that were considered, and discuss the decisions made at each step.
Usually, the final decision was based on the results of preliminary
experimentation with the identified options. We will describe two
parallel algorithms, sharing many characteristics. One, based on DFBnB,
we will refer to as PDFS. The other, based on A{*}, we will refer
to as PA{*}. For comparison purposes there is a variant of PA{*} which
includes a duplicate detection mechanism, which we will refer to as
PA{*}-DD. Algorithm \ref{alg:Pseudocode-PA*} gives pseudocode for
PA{*} and PA{*}-DD, and Algorithm \ref{alg:Pseudocode-PDFS} gives
pseudocode for PDFS.

\subsection{Work Decomposition}

\subsubsection{Unit of Work}

We start by identifying the parallelisable work in the algorithm.
Branch-and-bound search inherently comes with a division of work into
discrete and independent jobs (or tasks; in order to avoid confusion,
we will use the term job here). The natural unit of work is the expansion
of each individual state. Given two states, the children of each can
be constructed and evaluated simultaneously without any interaction
between processes.

\subsubsection{Central}

\begin{figure}
\begin{centering}
\includegraphics[width=5cm]{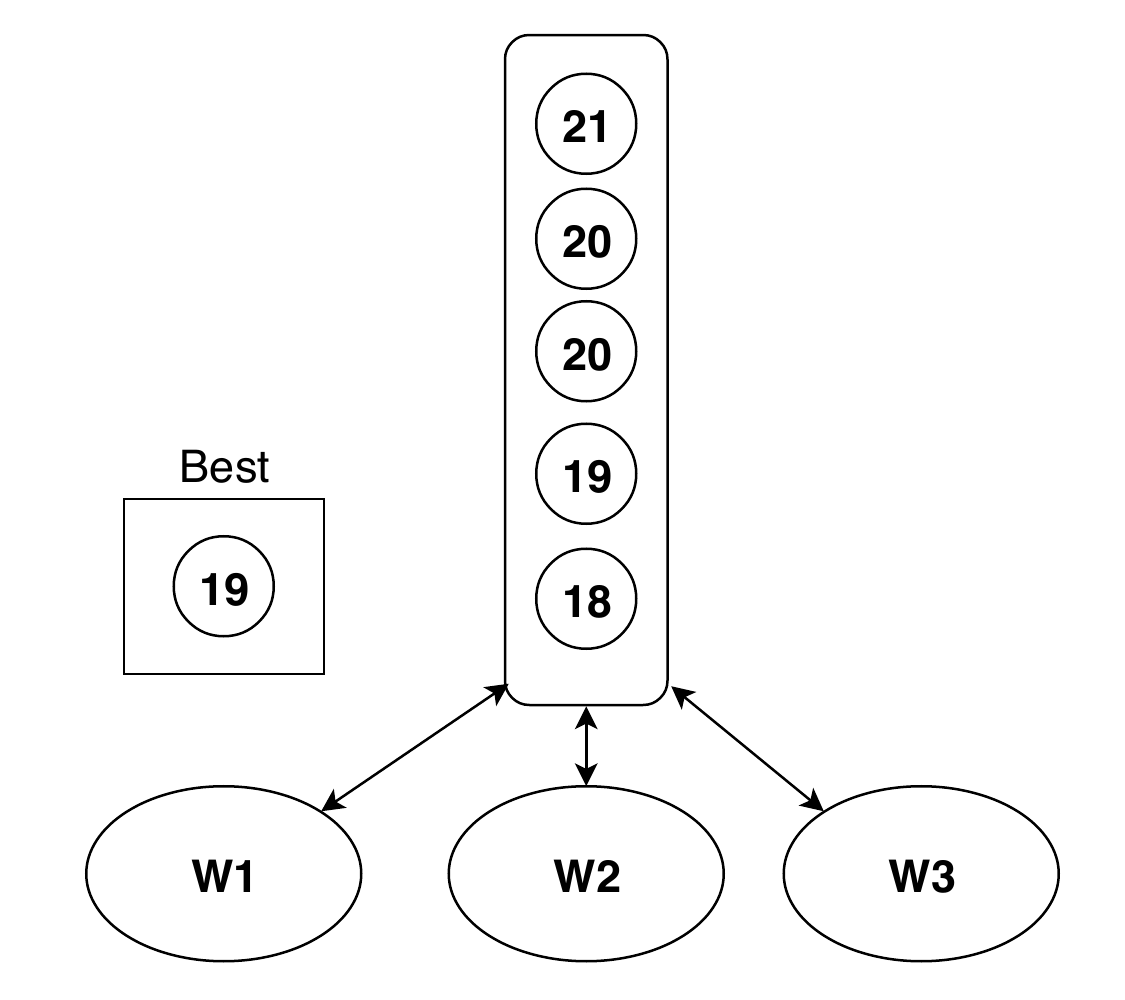}
\par\end{centering}
\caption{Workers sharing a central priority queue.\label{fig:Workers-central-PQ}}
\end{figure}

Given this, a natural method of parallelisation is a simple thread
pool, with a job queue from which a number of workers take states
and expand them, subsequently inserting the produced children back
into the queue. This method is visualised in Figure \ref{fig:Workers-central-PQ}.
It is usual to implement A{*} with a priority queue, so this does
not require an especially large change to the implementation of the
algorithm. However, use of this simple thread pool model is complicated
by the way in which A{*} requires the job queue to be used. The expansion
of a state will usually result in the creation of multiple child states,
which must then be added to the queue. This means that there must
be many more jobs inserted than retrieved, and the queue will inevitably
grow quickly. Since the best-first nature of A{*} necessitates a priority
queue, each of these insertions requires a non-trivial amount of work
- for a heap-based priority queue with current size $n$, this is
$O(log(n))$ time. Unlike a standard queue, insertions may cause changes
to any part of the data structure, meaning that parallel insertions
and retrievals may conflict with each other. The combination of these
factors makes operations on the priority queue a major source of contention
between workers. Investigation using Java's PriorityBlockingQueue
data structure, a standard binary heap with a global lock, showed
no speedup at all. Although a more complex data structure may have
allowed better performance, such as the more granularly locked pipelined
priority queue \citep{ioannou2007pipelined} or the lock-free skiplist
\citep{linden2013skiplist}, it is clear that an effective parallelisation
using this strategy is not so simple to implement as it may seem.
Previous research has also found that such an approach lead to performance
worse than sequential A{*}\citep{burns2010best}.

\subsubsection{Distributed}

\begin{figure}
\begin{centering}
\includegraphics[width=6cm]{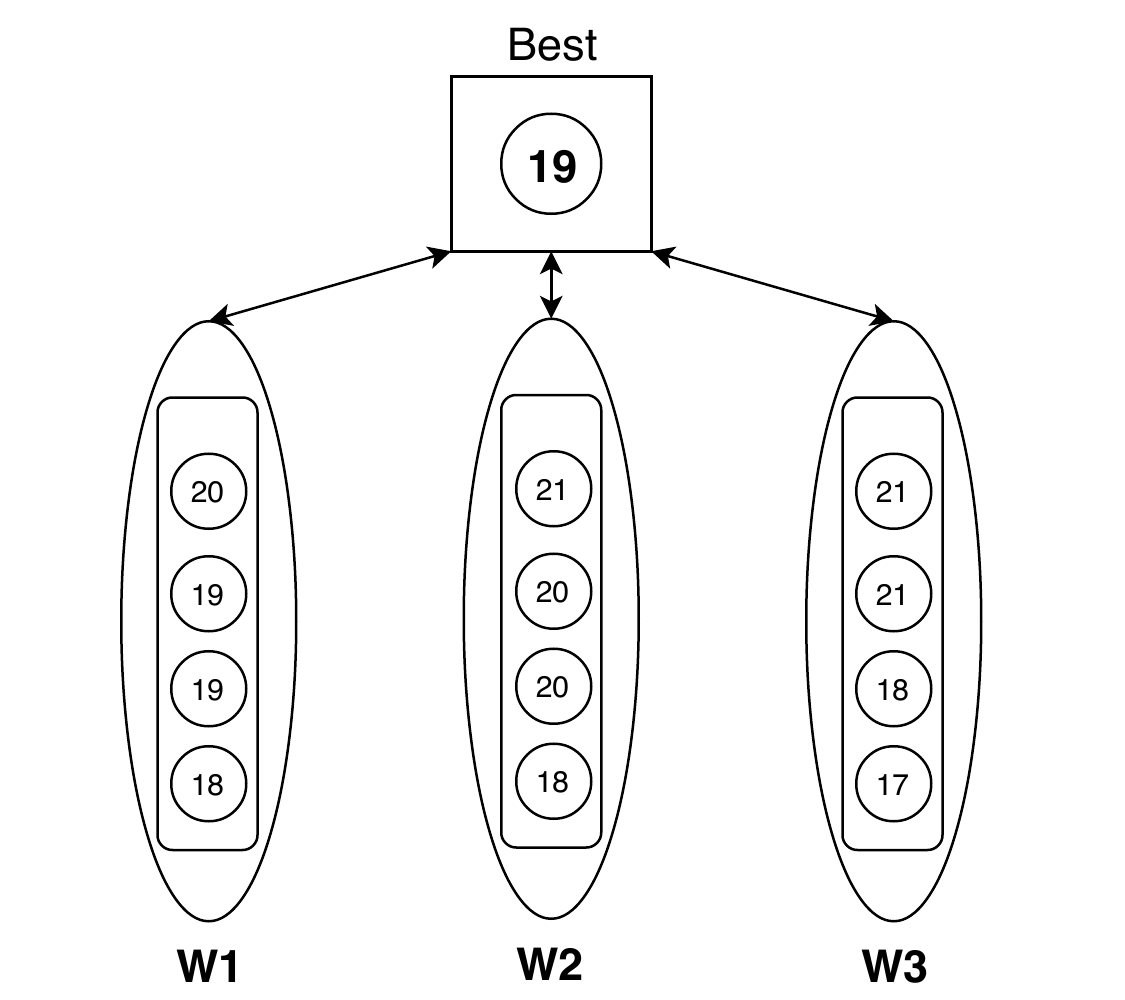}
\par\end{centering}
\caption{Distributed priority queues.\label{fig:Distributed-priority-queues.}}
\end{figure}

Instead of considering individual states as the basic unit of work,
we can instead have workers process entire sub-trees of the state-space,
as shown in Figure \ref{fig:Distributed-priority-queues.}. Each worker
has its own queue of states, which it can retrieve from as needed
and to which it will return all child states produced. This is a more
coarse-grained parallelisation strategy, requiring much less interaction
between workers. It does require, however, that states are initially
distributed between the workers. This can be achieved by beginning
with a stage in which a serial A{*} search is run until its queue
contains enough states to provide work for all workers. The states
can then be assigned to workers in round-robin fashion. Each worker
is then free to work on its own sub-tree, performing its own search
mostly in isolation. In the case of PA{*}, a heap-based priority queue
is used, while PDFS uses a linked-list-based deque. 

\subsection{Worker Synchronisation}

It is still necessary for some regular communication between workers
to occur in order for the overall search to be most efficient. The
problem with having sub-trees searched in isolation is that, since
the workers only have partial information about the state-space, the
pure best-first property of A{*} is no longer maintained. Although
each worker may select the best among the options available to it,
there is no guarantee that the selected state is anywhere close to
the true best currently available in the overall search. This means
that A{*}'s property of optimal efficiency no longer holds, and many
of the states expanded by a given worker may never have been touched
by a serial A{*} search. Similarly in a parallel DFBnB search if one
worker has already found a good solution but the other workers, without
knowledge of this, continue to expand states which could not lead
to a better solution, the time spent doing so is wasted. To mitigate
this, we must share information between workers, allowing them to
more often determine when expanding a particular state would be wasted
work. The more frequently information is shared, the closer to optimally
efficient we can be, but the greater the synchronisation overhead
will become. A simple way to do this, which works for both A{*} and
DFBnB, is to maintain a public record of the best solution found so
far in the overall search. Whenever a worker finds a complete solution,
it can compare it against this value, and update it if its new solution
is better. Workers will only examine states in their assigned sub-trees
with a lower $f$-value than their currently best known solution.
After examining a number of states (this is a tuneable parameter,
and we currently use onehundred thousand) they will check to see if
any other worker has discovered a better solution. With depth-first
search, solutions are expected to be discovered very often early on,
and less so as the optimal is approached. With A{*}'s best-first approach,
the opposite is expected, with no solutions at all found until near
the end of the search.

\subsubsection{Duplicate Detection}

The PA{*}-DD variant is defined by an additional data structure shared
between all the workers for the purposes of duplicate detection. The
data structure used is a hashmap: in our implementation, Java's ConcurrentHashMap,
a thread-safe hash map designed for high concurrency. When any worker
creates a new state, it will check if an identical state already exists
in the hashmap. If it does, the new state is a duplicate and is discarded.
If it does not, the new state is added to the hashmap and the algorithm
proceeds as normal.

\subsection{Load Balancing}

\begin{figure}
\begin{centering}
\includegraphics[width=6cm]{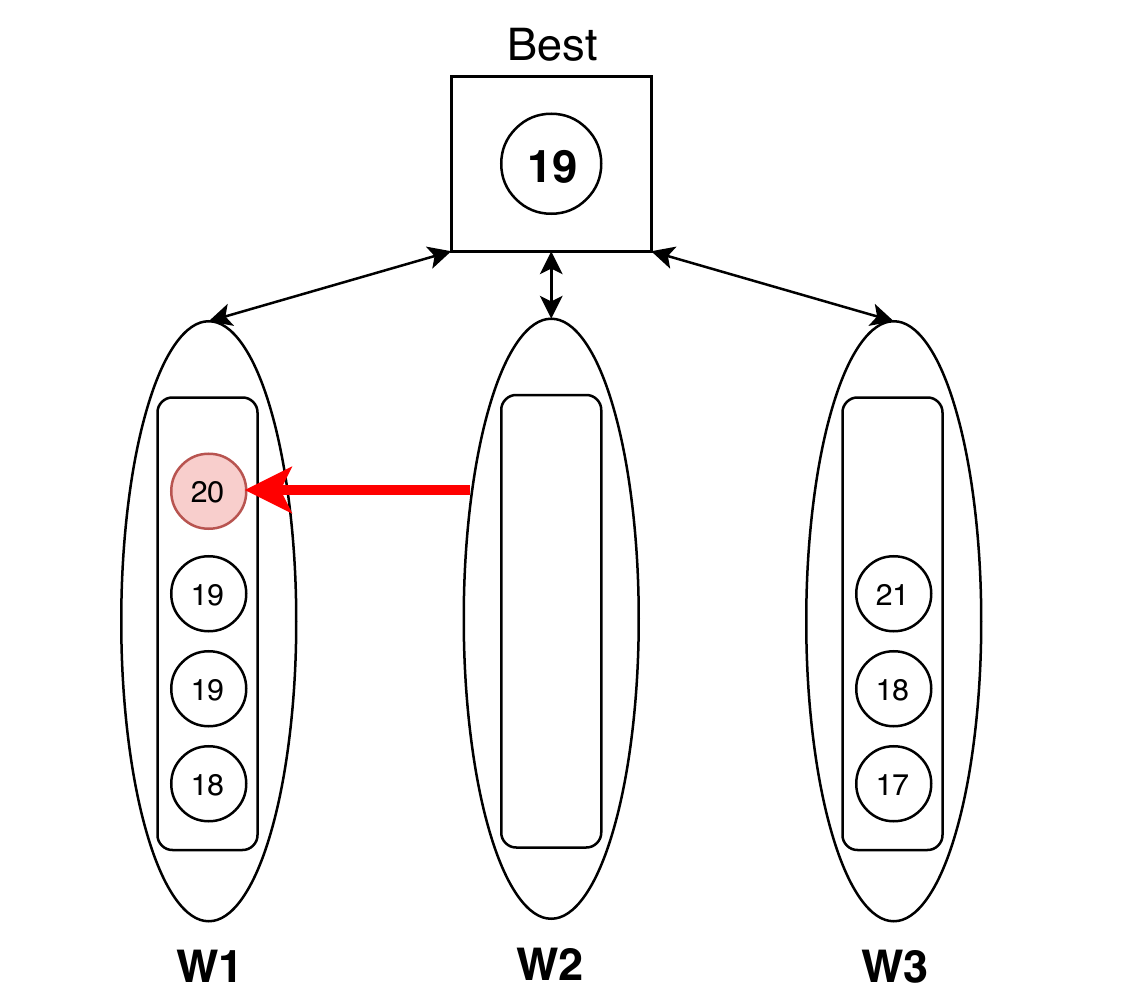}
\par\end{centering}
\caption{A worker with an empty queue steals from another.\label{fig:work-stealing-1}}
\end{figure}

Using entire sub-trees as units of work leads to another issue: it
is impossible to determine \emph{a priori} the amount of useful work
represented by a given sub-tree. Some root states may represent very
bad decisions, such that little or none of their descendants would
be considered by a serial A{*} search. Others may represent especially
good decisions, such that almost all states worth examining belong
to their sub-tree. This huge potential imbalance in work between sub-trees
means that some workers will finish much earlier than others, causing
speedup potential to be lost. To aid with load-balancing, we therefore
employ work-stealing \citep{blumofe1999scheduling}. When a worker
has exhausted all potentially useful work in its own sub-tree, it
visits another worker and takes a state from it, as shown in Figure
\ref{fig:work-stealing-1}. For DFBnB, a state is stolen from the
back end of the deque. This both minimises the chance of contention
between the thief and the victim, and maximises the total amount of
work stolen - since states at the tail of the deque are highest up
in the state-space, they lead to the largest subtrees. This will hopefully
ensure that the thief does not have to steal again soon after. For
A{*}, it is the current best state in the victim's priority queue
that is taken - meaning that it is the most likely state present to
lead to the optimal solution, and therefore most likely to represent
useful work. By contrast, stealing from the back of the queue could
yield a very low quality state which, if it is useful to examine it
at all, is relatively unlikely to lead to a significantly sized worthwhile
subtree. Stealing only a single state will minimise the impact of
synchronisation on the victimised worker. For both approaches, the
victims of stealing are selected randomly, which is the standard method
\citep{frigo1998implementation}. This tends to spread the burden
of victimisation uniformly and requires little communication between
threads for a decision to be made.

\subsection{Termination}

Another complication of this parallel branch-and-bound algorithm is
the difficulty of determining when a provably optimal solution has
been found, and the search can be terminated. In serial A{*}, the
best-first property ensures that the very first complete solution
found must necessarily be optimal. In this parallel version, the best-first
property is no longer maintained globally. Not only does this mean
that complete but non-optimal solutions may be discovered first, but
true optimal solutions may not immediately be able to be proven optimal.
In order to prove that a solution is optimal, both in A{*} and in
DFBnB, all states of lower $f$-value in the state-space must have
been examined and shown not to be complete solutions. Figures \ref{fig:Workers-central-PQ}
and \ref{fig:Distributed-priority-queues.} both depict a situation
in which a complete solution has been found (and stored as the best
known solution) while one or more states remain in the queue(s) which
have a lower f-value. For PDFS, this is part of standard operation.
If the algorithm being used is PA{*}, however, this must mean that
the loss of the strict best-first property led to one worker discovering
this complete solution while another worker was creating a child state
with a lower f-value. In both algorithms, a worker without any of
these lower-$f$-value states no longer has any useful work in its
queue, and it will attempt to steal a promising state from another
worker. An individual worker will know that it should terminate when
it has no more useful work in its queue, and there are no other workers
with work available to steal. The search will be finished only once
all workers have exhausted their queues of states with $f$-values
lower than the current best known solution. 

\begin{algorithm}[ph]
\begin{algorithm2e}[H]
	\KwIn{$S_i$, a set of states produced by sequential A*}
	\KwIn{\textbf{global} $W$, a set of workers}
	\KwIn{\textbf{global atomic} $bestSolution \gets \infty$}
	\KwIn{\textbf{global atomic} $idleWorkerCount \gets 0$}
	\KwOut{An optimal schedule}

	$priorityQueue \gets $ all $s \in S_i$, accessible as $W_i[PQ]$\;
	$hasWork \gets True$, accessible as $W_i[hasWork]$\;
	$localBestKnown \gets bestSolution$\;
	$syncCounter \gets 0$\;

	\If{$duplicateDetection = True$}{
		Insert all $s \in S_i$ into $hashmap$\;
	}
	\While(\tcp*[f]{at least one worker has work}){idleWorkerCount < |W|}{
		\While{$priorityQueue \neq \emptyset$}{
			$bestState \gets$ pop from $priorityQueue$\;
			
			$syncCounter \gets syncCounter + 1$\; 
			\If(\tcp*[h]{Periodic sync} ){$syncCounter = syncThreshold$}{
				$localBestKnown \gets bestSolution$\;
				$syncCounter \gets 0$\;
			}
			\eIf{$f(bestState)< f(localBestKnown)$}{
				\If{$complete(bestState)$}{
					Test-and-set $bestSolution \gets bestState$\;
				}
				\For{$c \in children(bestState)$}{
					\If{$f(c)<f(localBestKnown)$}{
						\eIf(\tcp*[f]{DD variant}){$duplicateDetection=True$}{
							\If{$c \notin hashmap$}{
								Insert $c$ into $hashmap$\;
								Insert $c$ into $priorityQueue$ with priority $f(c)$\;
							}
						}{
							Insert $c$ into $priorityQueue$ with priority $f(c)$\;
						}
					}
				}
			}(\tcp*[h]{No more useful work}){
				$priorityQueue \gets \emptyset$\;
			}
		}
		$hasWork \gets False$\;
		Test-and-set $idleWorkerCount \gets idleWorkerCount + 1$\;
		
		$victim \gets$ a random number from $0$ to $|W|$\;
		\If(\tcp*[f]{Work Stealing} ){$W_{victim}[hasWork]=True$}{
			$stolenState \gets$ pop \textbf{head} from $W_{victim}[PQ]$\;
			Insert $stolenState$ into $priorityQueue$\;
			$hasWork \gets True$\;
			Test-and-set $idleWorkerCount \gets idleWorkerCount - 1$\;
		}
	}
	\Return{$bestSolution$}
\end{algorithm2e}

\caption{Pseudocode for a single PA{*} worker.\label{alg:Pseudocode-PA*}}
\end{algorithm}
\begin{algorithm}[ph]
\begin{algorithm2e}[H]
	\KwIn{$S_i$, a set of states produced by sequential A*}
	\KwIn{\textbf{global} $W$, a set of workers}
	
	\KwIn{\textbf{global atomic} $bestSolution \gets \infty$}
	\KwIn{\textbf{global atomic} $idleWorkerCount \gets 0$}
	\KwOut{An optimal schedule}

	$stack \gets \emptyset$, accessible as $W_i[stack]$\;
	$hasWork \gets True$, accessible as $W_i[hasWork]$\;
	$localBestKnown \gets bestSolution$\;
	$syncCounter \gets 0$\;
	Push all $s \in S_i$ onto $stack$\;
	\While(\tcp*[f]{at least one worker has work}){idleWorkerCount < |W|}{
		\While{$stack \neq \emptyset$}{
			$currentState \gets$ pop from $stack$\;
			
			$syncCounter \gets syncCounter + 1$\; 
			\If(\tcp*[h]{Periodic sync} ){$syncCounter = syncThreshold$}{
				$localBestKnown \gets bestSolution$\;
				$syncCounter \gets 0$\;
			}
			\If{$f(currentState)< f(localBestKnown)$}{
				\If{$complete(currentState)$}{
					Test-and-set $bestSolution \gets currentState$\;
				}
				\For{$c \in children(bestState)$}{
					\If{$f(c)<f(localBestKnown)$}{
							Push $c$ onto $stack$\;
					}
				}
			}
		}
		$hasWork \gets False$\;
		Test-and-set $idleWorkerCount \gets idleWorkerCount + 1$\;

		$victim \gets$ a random number from $0$ to $|W|$\;
		\If(\tcp*[f]{Work Stealing} ){$W_{victim}[hasWork]=True$}{
			$stolenState \gets$ pop \textbf{tail}  from $W_{victim}[stack]$\;
			Push $stolenState$ onto $stack$\;
			$hasWork \gets True$\;
			Test-and-set $idleWorkerCount \gets idleWorkerCount - 1$\;
		}
	}
	\Return{$bestSolution$}
\end{algorithm2e}

\caption{Pseudocode for a single PDFS worker.\label{alg:Pseudocode-PDFS}}
\end{algorithm}

\subsection{Evaluation}

To experimentally evaluate the hypothesis that AO would allow better
performance for parallel search algorithms, we performed parallel
searches on a set of task graphs using the proposed parallel algorithms
of A{*} and DFBnB. Task graphs were chosen that differed by graph
structure and communication-to-computation ratio (CCR). From the larger
dataset described in Table \ref{tab:Range-of-task-1}, the group of
graphs with 21 tasks were selected. This is a set of 340 task graphs.
We attempted to find an optimal schedule with 4 processors, for both
state-space models, using 1, 2, 4, 8, 16 and 24 worker threads, once
each for PDFS and PA{*}. For each trial using ELS with PA{*}, an additional
trial using PA{*}-DD was added. This gave a total of 10,200 trials.
The algorithms were implemented in the Java programming language,
based on sequential implementations of both ELS and AO \citep{2019arXiv190106899O}.
All tests were run on a Linux machine (Ubuntu 16.04) with 4 Intel
Xeon E7-4830 v3 @2.1GHz processors. This system has a total of 48
cores. Each trial wasallowed a time limit of 2 minutes to complete.
For all tests, the JVM (Java HotSpot 25.91) was given a maximum heap
size of 96 GB. Each search was started in a new JVM instance, to remove
the possibility of previous searches impacting them through garbage
collection and JIT compilation. 

\begin{figure}
\begin{centering}
\includegraphics[width=12cm]{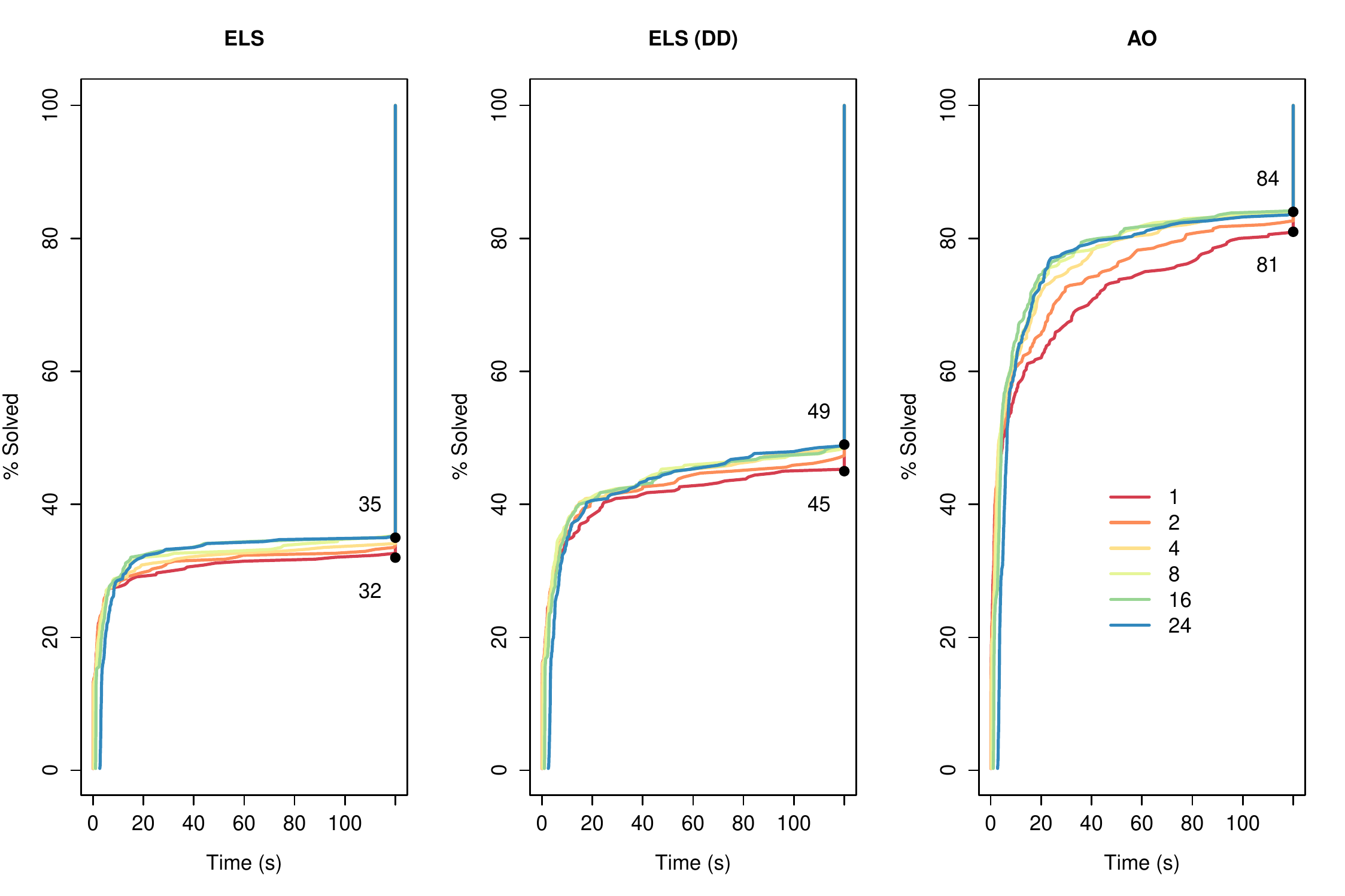}
\par\end{centering}
\caption{Performance profiles for PA{*} and PA{*}-DD\label{fig:Performance-profiles-parallelA*}}
\end{figure}
Figure \ref{fig:Performance-profiles-parallelA*} shows performance
profiles (as described in Section \ref{sec:Depth-First-Branch-and})
demonstrating how performance with PA{*} varied across the range of
threads used. It is clear from these profiles that the AO model consistently
has the best absolute performance, solving the most instances after
any given time, while ELS solves significantly less instances without
duplicate detection. This confirms our expectations regarding the
state-space models and the importance of duplication detection.

All three variants show some amount of performance increase, scaling
with the number of threads. By the time limit of 2 minutes, the multithreaded
variants solved 3-4\% more instances than in the sequential case.
However, with AO, we can see clearly that at earlier times there are
greater differences between the curves, suggesting that parallelisation
is in fact solving instances much quicker. The flattening of all the
curves tells us that fewer and fewer problem instances are solved
as time goes on, which is the expected effect in an exponential state-space
search.This effect allows the sequential algorithm (and parallelisation
with lower number of threads) to ``catch up''.

\begin{figure}
\begin{centering}
\includegraphics[width=12cm]{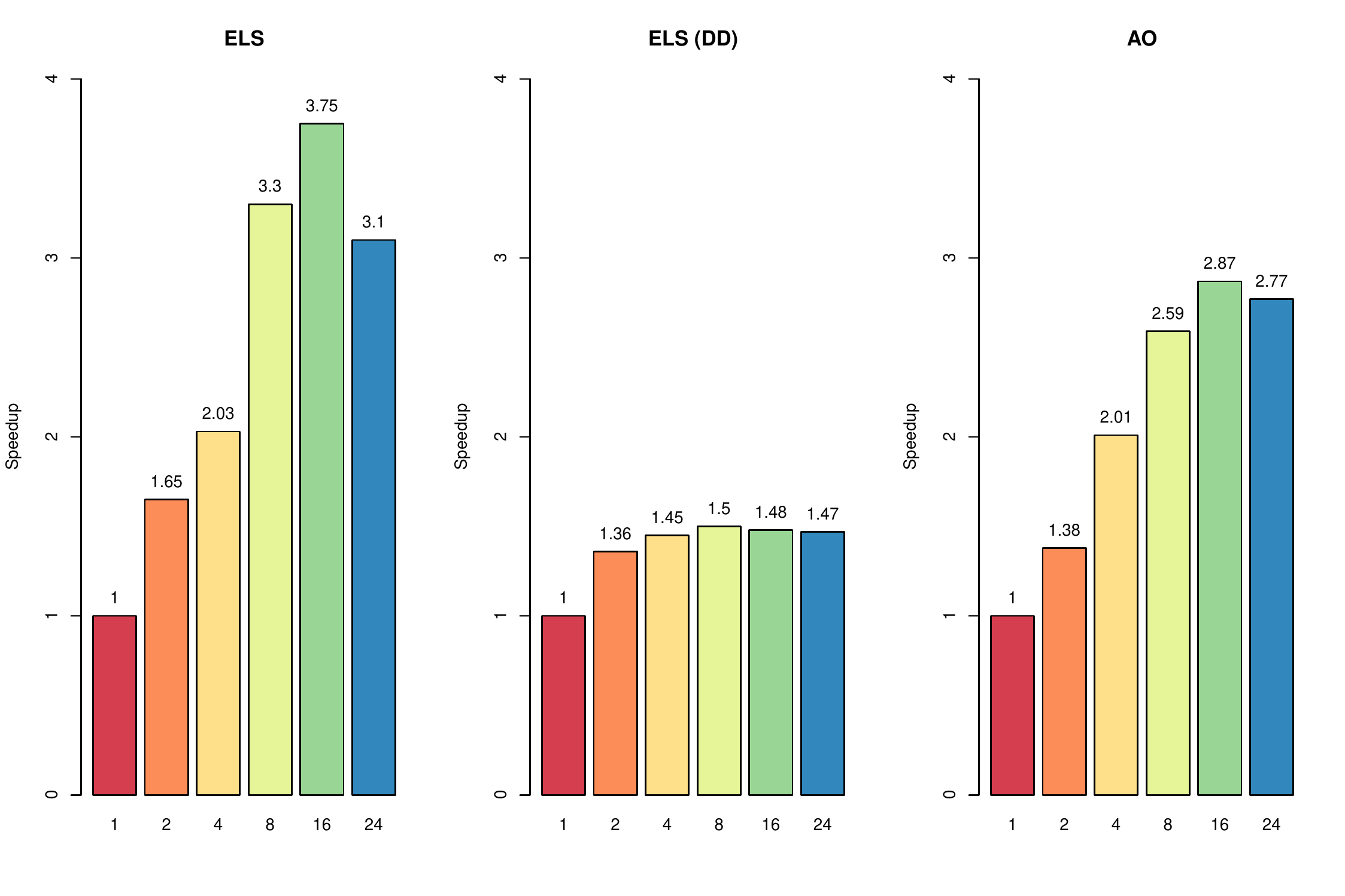}
\par\end{centering}
\caption{Speedup of PA{*} at 60 seconds over different number of threads.\label{fig:Speedup-of-parallelA*}}
\end{figure}
In order to better analyse the scaling of the parallel algorithm,
and considering the flattening effect just discussed, we calculate
the speedup after $x$ seconds of the parallel algorithm with $n$
threads as follows: first, we find the number of problem instances
solved by the sequential algorithm within $x$ seconds,$\text{seqnum}(x)$.
Then, we find the time taken for the parallel algorithm with $n$
threads to solve $\text{seqnum}(x)$ problem instances, $\text{partime}_{n}(\text{seqnum}(x))$.
Finally, the speedup is defined as $\text{speedup}(n,x)=x/(\text{partime}_{n}(\text{seqnum}(x)))$.
Figure \ref{fig:Speedup-of-parallelA*} shows the result of this calculation
across our dataset with $x$ at 60 seconds. By this metric we see
that both AO and ELS with PA{*} show consistent scaling as the number
of threads is increased. The exception is a decrease in performance
between 16 and 24 threads, likely a sign that synchronisation between
threads produced too much overhead at that level of parallelisation.
While it is relatively weak, it is very encouraging that scaling is
seen across a large number of threads, demonstrating the potential
of the method. A reason for the lack of better scaling might be the
use of the Java language and standard concurrent data structures.ELS
with PA{*}-DD does not show consistent scaling. This corresponds with
our hypothesis that the use of a duplicate detection mechanism would
negatively impact the benefit gained from a parallel algorithm.

\begin{figure}
\begin{centering}
\includegraphics[width=12.5cm]{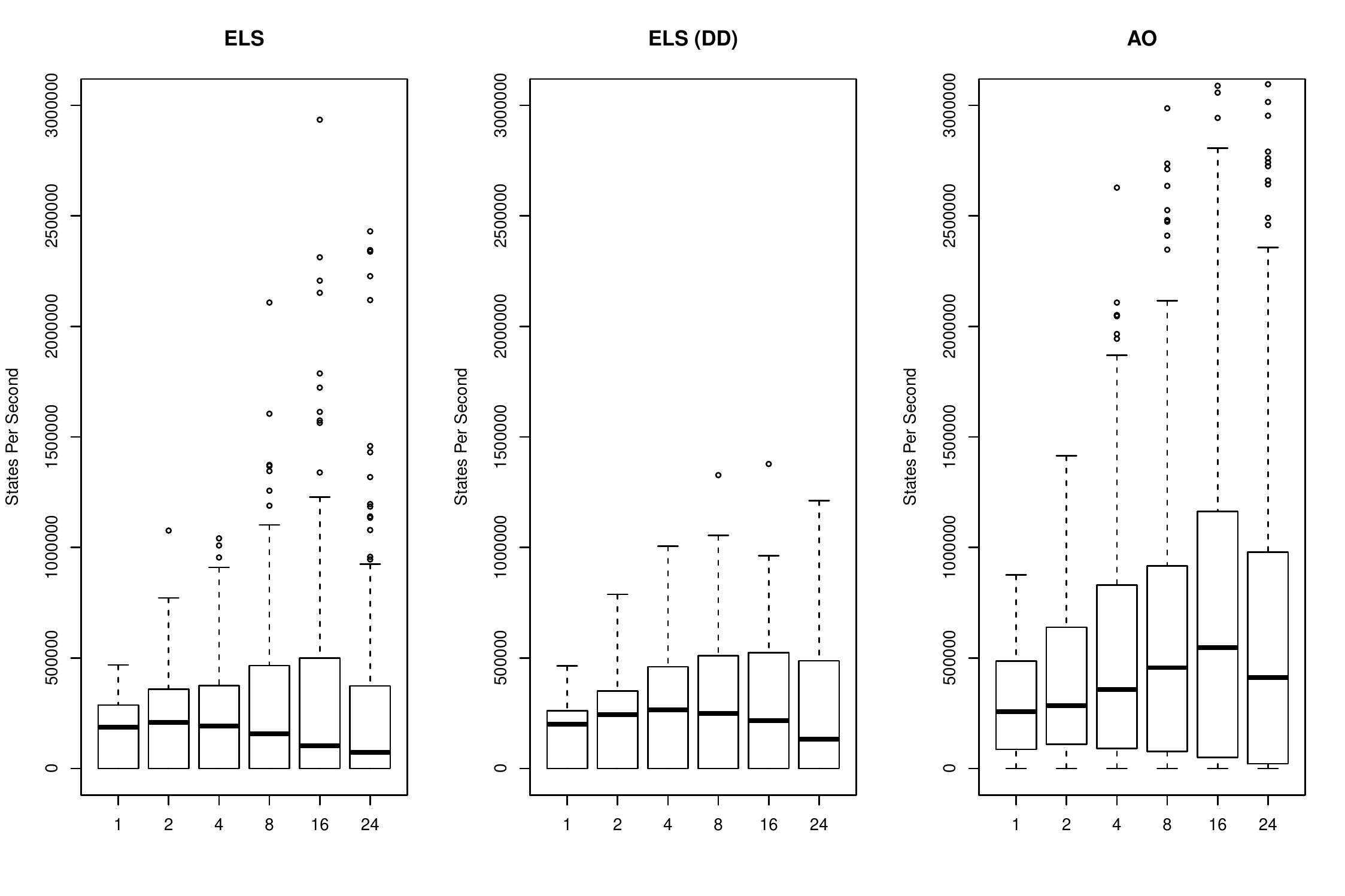}
\par\end{centering}
\caption{States examined per second by PA{*}.\label{fig:States-per-second-parallelA*}}
\end{figure}
Another way to measure the benefit gained from the parallel algorithm
is to examine how quickly states are examined with varying numbers
of threads. This can be considered as a more direct measurement of
how much work is being done by the search algorithm. Note that this
metric cannot be directly correlated with the actual amount of problem
instances which are solved. One reason for this is that not all of
the additional work performed by the parallel algorithm will be ``useful''
work. It is likely that many states will be examined that would not
have been touched by the sequential algorithm, meaning that an increase
in total work performed will not translate directly to an increase
in performance. Figure \ref{fig:States-per-second-parallelA*} contains
box plots showing how the distribution of states per second changes
with the number of threads, for each model. Here we see that for AO
the number of states per second tends to increase consistently as
the degree of parallelisation increases. ELS does not display similar
trends with either PA{*} or PA{*}-DD.

\begin{figure}
\begin{centering}
\includegraphics[width=0.6\textwidth]{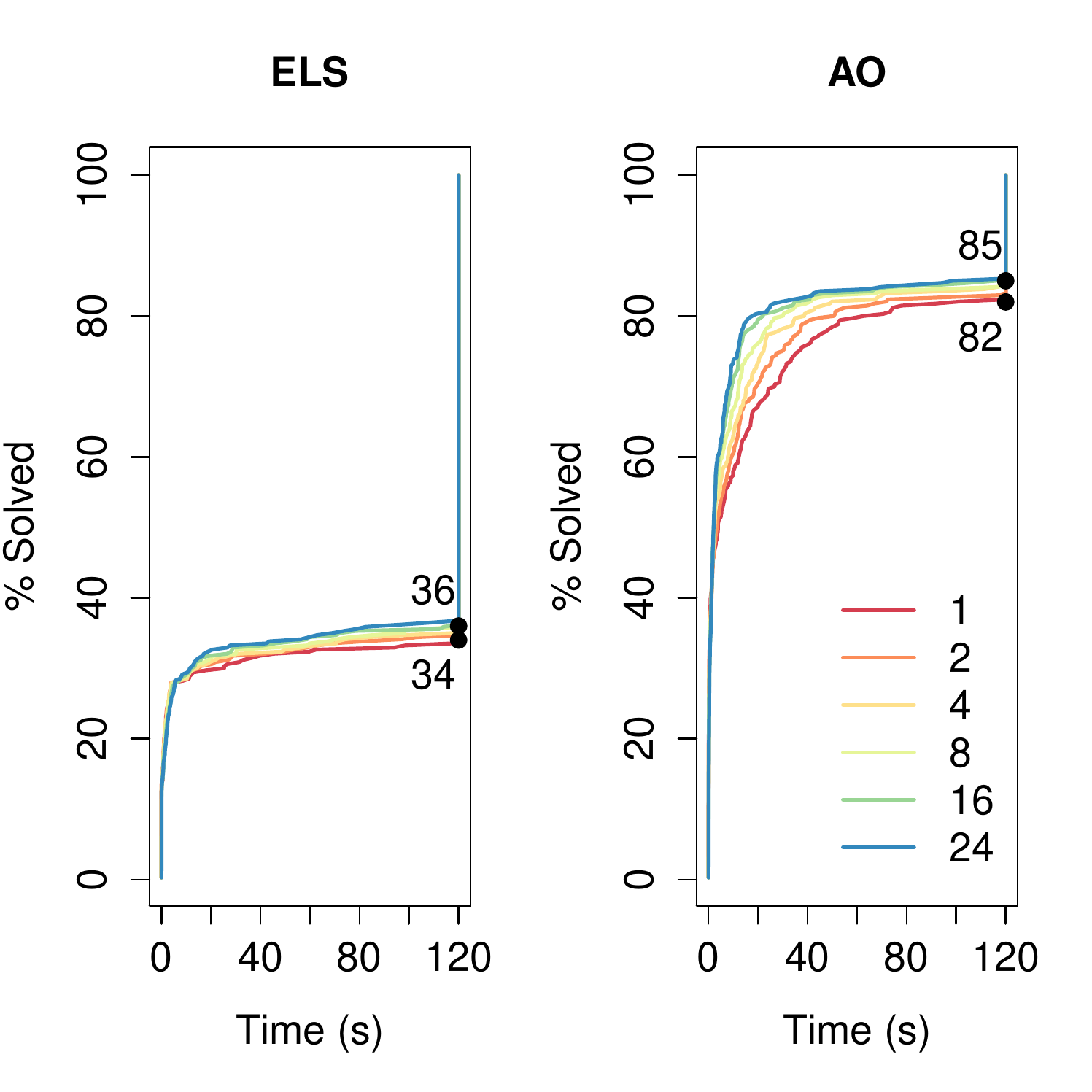}
\par\end{centering}
\caption{Performance profiles for PDFS.\label{fig:Performance-profiles-PDFS}}
\end{figure}
Figure \ref{fig:Performance-profiles-PDFS} shows performance profiles
demonstrating how performance with PDFS varied across the range of
threads used. Similar to the profiles for PA{*}, both models see an
increase of 2-3\% of problem instances solved with the use of parallelisation,
after 120 seconds have elapsed. However, the absolute performance
of AO is much better than that of ELS: AO solves more than twice as
many problem instances within the time limit. Interestingly, DFBnB
shows itself to be competitive with A{*} when used with the AO model,
with almost the same proportion of problem instances solved. This
is remarkable, as under the ELS model DFBnB is not performance competitive
with A{*} \citep{venugopalan2016memory}. This can also be observed
when comparing Figure~\ref{fig:Performance-profiles-parallelA*}
with Figure~\ref{fig:Performance-profiles-PDFS}, where we see that
the single threaded PA{*}-DD under the ELS model (ELS(DD)) performs
significantly better than any PDFS under ELS. In general, it is clear
that performance with the AO model has scaled with the number of threads
used in a similar manner as with PA{*}, but it is difficult to distinguish
a trend in the ELS results. 

\begin{figure}
\begin{centering}
\includegraphics[width=0.6\textwidth]{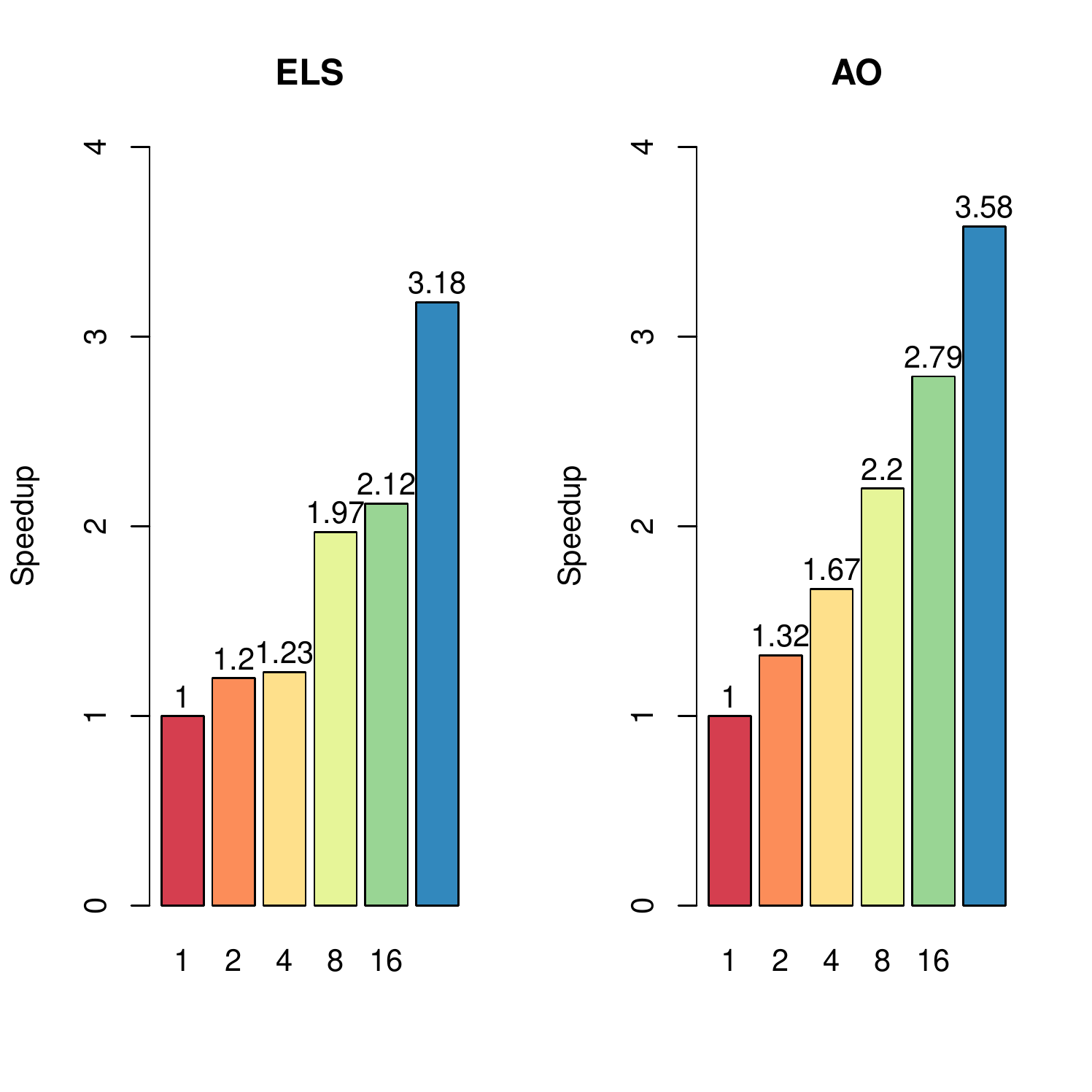}
\par\end{centering}
\caption{Speedup of PDFS at 60 seconds.\label{fig:Speedup-of-PDFS}}
\end{figure}
Figure \ref{fig:Performance-profiles-PDFS} shows speedup calculated
at 60 seconds for PDFS. Both ELS and AO show similar trends with speedup
scaling consistently with the number of threads used. This view shows
us that performance scales weakly, but consistently with the number
of threads used for both state-space models, with ELS reaching a maximum
speedup of 3.18 and AO reaching a maximum speedup of 3.58. Note that,
unlike with PA{*}, the performance is not degraded when moving from
16 to 24 threads. This suggests that PDFS may continue to scale with
higher numbers of threads before synchronisation overhead becomes
too much.

\section{Conclusions\label{sec:Conclusions}}

A new state-space model for optimal task scheduling was recently proposed,
in which duplicate states are not produced. This state-space model
is known as AO, or Allocation-Ordering. We expected that this would
be advantageous to a wide variety of branch-and-bound methods, in
particular depth-first branch-and-bound and parallel algorithms. We
have therefore investigated and proposed DFBnB and parallel search
algorithms for optimal task scheduling. This includes a parallel algorithm
based on A{*} (PA{*}) and one based on DFBnB (PDFS).

It was hypothesised that memory-limited search, such as with DFBnB,
would produce better results with AO than with ELS, an older state-space
model with many duplicate states. Our extensive experimental evaluation
showed that AO was greatly superior to ELS when used with DFBnB, solving
between 50\% and 90\% more problem instances within the time limit.

It was also considered likely that the lack of duplicates would allow
parallel branch-and-bound to scale better with AO than with ELS, as
the lack of data structures associated with duplicate detection would
mean lower levels of synchronisation. The experimental evaluation
of our proposed algorithms demonstrated that AO allowed the performance
of parallel A{*} and DFBnB algorithms to scale better with the number
of threads used, with problems that took the sequential algorithm
60 seconds being solved up to 3.58 times faster. While the scaling
was not very strong, it was encouraging that it was consistent over
a good number of threads. 

A combination of comparable absolute performance, more consistent
scaling in parallel, and the memory-limited nature of the DFBnB algorithm
suggest that PDFS with the AO model is the best candidate method for
optimal task scheduling with state-space search. This is a new result,
as so far A{*} (using the ELS model) was superior to DFBnB.

\bibliographystyle{ieeetr}
\bibliography{journalref}

\end{document}